\documentclass{emulateapj}
\usepackage{apjfonts}

\usepackage{amsmath}
\usepackage{graphicx}

\usepackage{comment}
 \usepackage{color}           
 \definecolor{DarkGreen}{rgb}{0.0,0.45,0.0}  

\newcommand{\sat}[1]{\it\uppercase{#1}\rm}
\newcommand{\fig}[1]{Fig.~\ref{#1}}

\newcommand{\sect}[1]{Section~\ref{#1}}

\newcommand{\accel}[1]{#1 m~s${}^{-2}$}

\begin{document}
\title{A Prominence Eruption Driven by Flux Feeding from Chromospheric Fibrils}
\author{Quanhao Zhang, Rui Liu, Yuming Wang, Chenglong Shen, Kai Liu, Jiajia Liu, and S. Wang}
\affil{CAS Key Laboratory of Geospace Environment, Department of Geophysics and Planetary Sciences, University of Science and Technology of China, Hefei 230026, China; rliu@ustc.edu.cn}

\begin{abstract}

We present multi-wavelength observations of a prominence eruption originating from a quadrupolar field configuration, in which the prominence was embedded in a side-arcade. Within the two-day period prior to its eruption on 2012 October 22, the prominence was perturbed three times by chromospheric fibrils underneath, which rose upward, became brightened, and merged into the prominence, resulting in horizontal flows along the prominence axis, suggesting that the fluxes carried by the fibrils were incorporated into the magnetic field of the prominence. These perturbations caused the prominence to oscillate and to rise faster than before. The absence of intense heating within the first two hours after the onset of the prominence eruption, which followed an exponential increase in height, indicates that ideal instability played a crucial role. The eruption involved interactions with the other side-arcade, leading up to a twin coronal mass ejection, which was accompanied by transient surface brightenings in the central arcade, followed by transient dimmings and brightenings in the two side-arcades. We suggest that flux feeding from chromospheric fibrils might be an important mechanism to trigger coronal eruptions.
\end{abstract}


\section{Introduction}
Solar \emph{prominences} are thread-like clouds consisting of relatively cool and dense magnetized plasma, suspended in the hot and tenuous corona. Known also as \emph{filaments} (used interchangeably with prominences hereafter), they appear as dark features along the polarity inversion line (PIL) when viewed on the disk, typically in H$\alpha$ filtergrams. Twisted/sheared magnetic field plays a crucial role in the equilibrium and dynamic evolution of prominences \citep[][and references therein]{Mackay2010a}. The remarkably stable equilibrium achieved by prominences and their sudden eruptions pose a great challenge for our understanding of the physics governing the destabilizing of the solar corona. Prominence eruptions have close association with flares and coronal mass ejections \citep[CMEs; see the reviews by][]{Low1996a, Chen2011a}. The three eruptive phenomena are hence suggested to be different manifestations of a single physical process, which involves the large-scale disruption and restructuring of the coronal magnetic field \citep{forbes00,pf02,lsb03}.
\par 
It is widely accepted that the corona is energized by photospheric or sub-photospheric activities, such as shearing motions near PILs \citep[e.g.,][]{kusano02, moon02}, emerging magnetic flux \citep[e.g.,][]{Feynman1995a}, and flux cancellation \citep[e.g.,][]{livi89}. But it is still under debate how exactly coronal eruptions are triggered. A large number of mechanisms have been proposed, which include, but are not limited to, tether-cutting reconnection in a sheared arcade \citep{Moore2001a}, breakout reconnection at a magnetic null point in a quadrupolar configuration \citep{Antiochos1999a}, flux emergence \citep{Chen2000a}, catastrophic loss of equilibrium \citep{fi91} through either photospheric flux cancellation \citep[e.g.,][]{linker03} or an artificial increase in either the poloidal or the axial flux of a flux rope \citep[e.g.,][]{su11}, and ideal MHD instabilities such as the helical kink instability \citep[e.g.,][]{fan05} and the torus instability \citep{kt06}. Recently, \citet{liu12} studied a ``double-decker'' filament, which was composed of two branches separated in height. They found that prior to the eruption of the upper branch, multiple filament threads within the lower branch brightened up, rose upward, and merged into the upper branch. This transfer of magnetic flux and current to the upper branch is suggested to be the key mechanism responsible for its loss of equilibrium by reaching the limiting flux that can be stably held down by the overlying field \citep{su11} or by reaching the threshold of the torus instability \citep{kt06}. 

\par
In this paper, we present the observation of a similar transfer of magnetic flux to a prominence through multiple rising `mini-prominences' originally located on the surface.  In the sections that follows, we investigate the evolution and eruption of the prominence, which was embedded in the side-arcade of a quadrupolar field configuration (\sect{sec:Observations}). We argue that these mini-prominences are of the same nature as chromospheric fibrils (\S\ref{sec:nature}), and then discuss the relevant mechanisms for the prominence eruption (\S\ref{sec:instability} and \S\ref{sec:reconnection}), which was apparently coupled to the eruption of the other side-arcade in the quadrupolar field.

\section{Observations and Analysis}
\label{sec:Observations}

\begin{figure*}
\includegraphics[width=\textwidth]{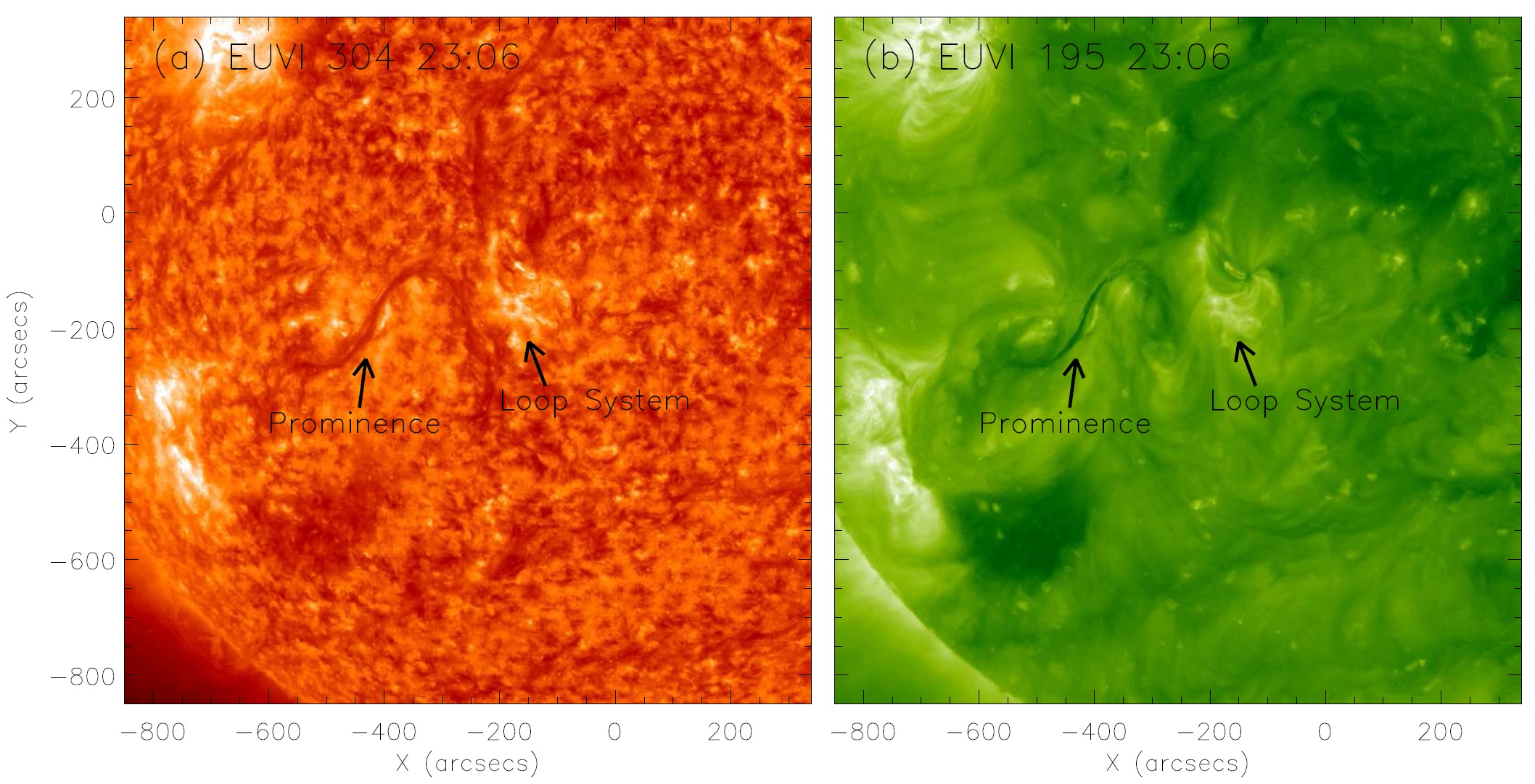}
\caption{The prominence and the loop system from STA's perspective in 304 and 195~{\AA}}
\label{fig:orig}
\end{figure*}
\subsection{Instruments}
The prominence was observed on the west limb in EUV by the Atmospheric Imaging Assembly \citep[AIA;][]{lemen12} onboard the Solar Dynamics Observatory \citep[SDO;][]{pesnell12}, and in H$\alpha$ by the Kanzelh\"{o}he Solar Observatory (KSO). Images taken by the Extreme Ultraviolet Imager \citep[EUVI;][]{wuelser04}) of the Sun Earth Connection Coronal and Heliospheric Investigator \citep[SECCHI;][]{howard08} imaging package onboard the Solar TErrestrial RElations Observatory \citep[STEREO;][]{kaiser08}) were utilized to provide a different perspective of this prominence, which appeared as a filament in the field of view (FOV) of STEREO's `Ahead' spacecraft (hereafter STA). The CME resulting from this eruption was observed by the Large Angle and Spectrometric Coronagraph \citep[LASCO;][]{brueckner95} onboard the Solar and Heliospheric Observatory (SOHO). Magnetograms obtained by the Helioseismic and Magnetic Imager \citep[HMI;][]{schou12a,schou12b} onboard SDO provide the magnetic context of the eruption's source region.

\subsection{Eruptive Process}
\label{sec:erupt}

\begin{figure*}
\includegraphics[width=\textwidth]{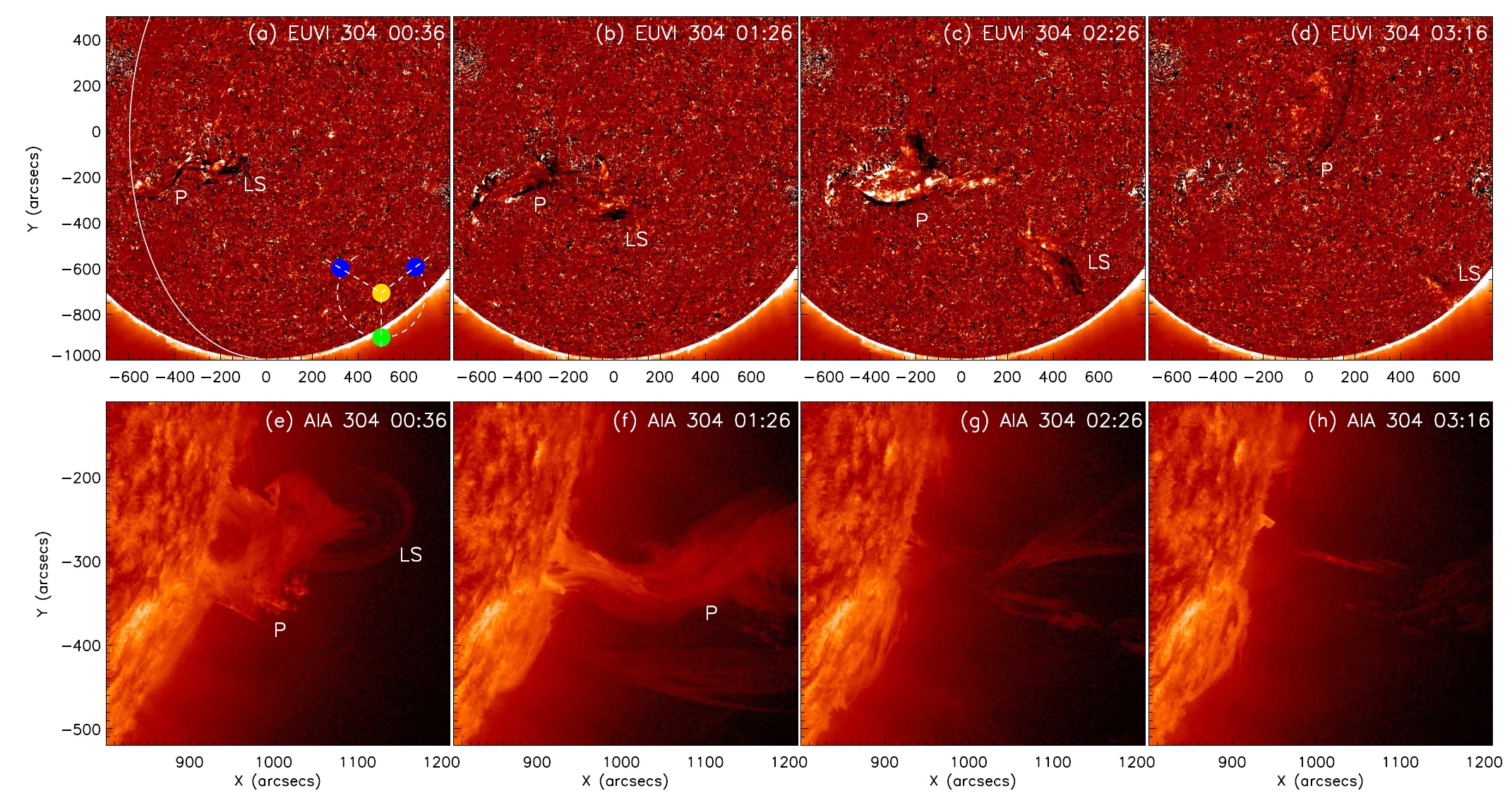}
\caption{Eruption of the prominence observed by STA (top) and SDO (bottom). Top panels show running-difference images in EUVI 304~{\AA}, and bottom panels the corresponding original images in AIA 304~{\AA} taken at approximately the same time. The loop system is marked as `LS', and the prominence as `P'. In Panel (a), the curve denotes the solar limb as seen by SDO, and the inset plots the positions of the STEREO spacecrafts (blue dots) relative to the Sun (yellow dot) and Earth (green dot) in the plane of the Earth's orbit, with STA ahead of, and STB behind, the Earth. An animation of AIA and EUVI 304~{\AA} images is available online.}
\label{fig:total_erupt}
\end{figure*}

The prominence as observed by SDO erupted from the west limb on the early 2012 October 22. It was observed simultaneously by STA/EUVI as a filament located in the southeast quadrant of the disk, not far from the disk center (\fig{fig:orig}). \fig{fig:total_erupt} shows the eruptive process in 304~{\AA} as monitored by both satellites. From SDO's perspective, a loop system (hereafter LS) was overlying the prominence (labeled `P') in projection (\fig{fig:total_erupt}(e)); from STA's perspective LS was located to the west of the prominence (Fig.~\ref{fig:orig}). The eruption started as early as 23:49 UT on 2012 October 21 (see \S\ref{subsec:height}). The LS erupted southwestward and left the disk at about 03:16 UT  (STA's perspective; \fig{fig:total_erupt}(d)), whereas the prominence erupted northwestward, and was still projected onto the disk center by the same time. This does not necessarily imply a difference in their propagation speeds, but due probably to LS's faster expansion, as can be seen from the bottom panels of \fig{fig:total_erupt}. It is noteworthy that the prominence was apparently writhed at the onset of the eruption (SDO's perspective; \fig{fig:total_erupt}(e)), taking on a projected forward S-shape on the disk (STA's perspective; \fig{fig:orig}). During the eruption, the prominence underwent a clockwise rotation of its axis, and consequently the S-shape was apparently straightened (\fig{fig:total_erupt}(b), see also the accompanying animation). This is opposite to the conversion of magnetic twist into writhe, in which case a counterclockwise rotation is expected if the flux rope assumes a forward S-shape \citep{green07,torok10}. 

\begin{figure*}
\includegraphics[width=\textwidth]{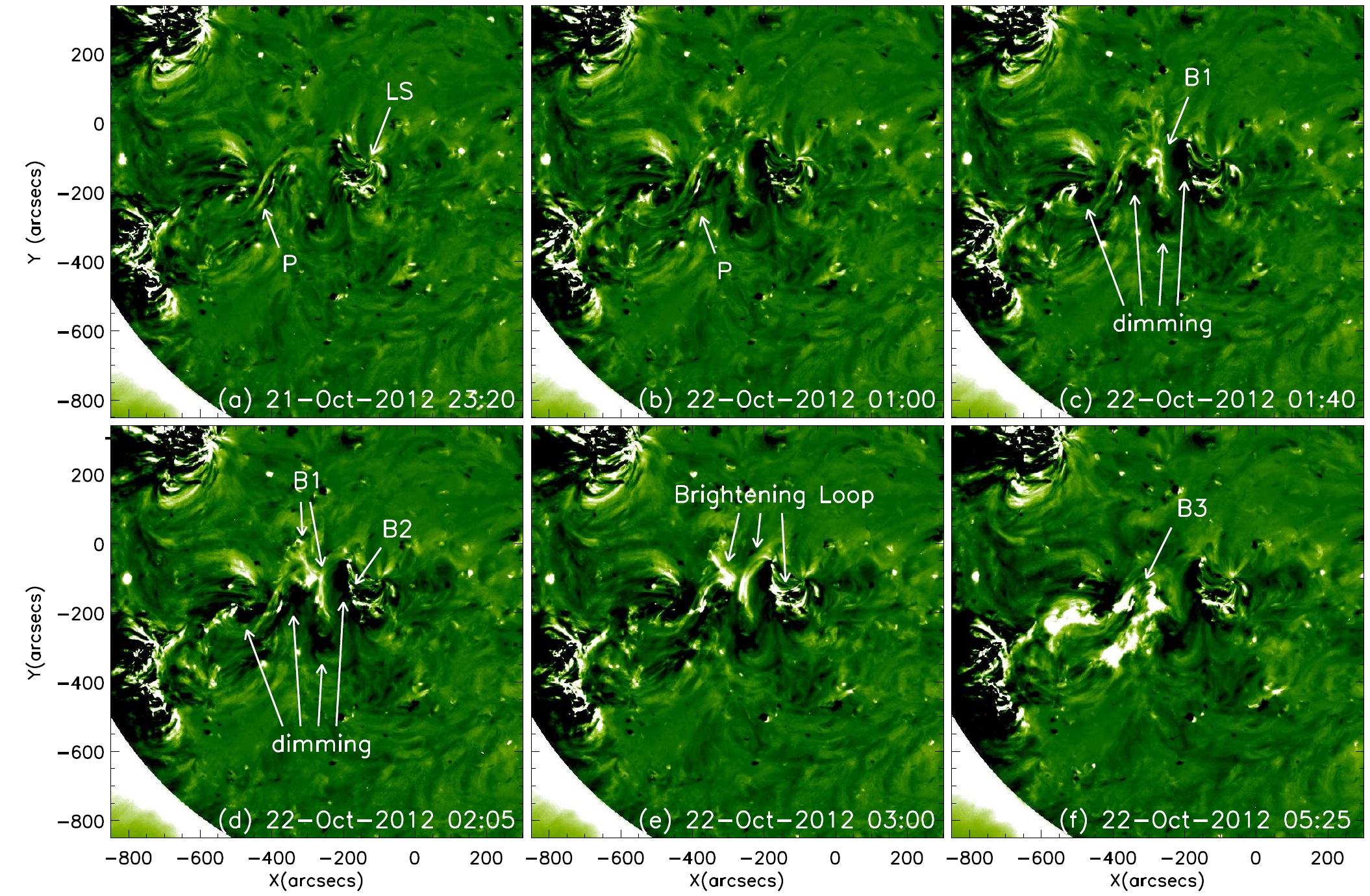}
\caption{Eruptions and the atmospheric response as observed in STA/EUVI 195~{\AA}. All images are subtracted by the `base' image taken at 09:40:30 UT on 2012 October 21. Panel (a) shows an pre-eruption image, with the source regions of the LS and the prominence (P) marked by arrows. The eruptive process is featured in Panels (b)-(f) with transient brightenings (labeled B1-B3) and dimmings marked by arrows. Brightening loops evolving from B1 and B2 are labeled T1 and T2, respectively, in Panel (e).} An animation of both original and running-difference 195~{\AA} images is available online.
\label{fig:erupt_195}
\end{figure*}

The sequence of the eruption is best demonstrated by EUVI 195~{\AA} base-difference images in Fig.~\ref{fig:erupt_195}, though the LS was not quite visible on the disk in 195~{\AA}. A sequence of reconnection events is characterized by successive brightenings at the surface. The 1st episode of brightening (B1) occurred in the central region between the prominence and the LS at about 01:40 UT (Fig.~\ref{fig:erupt_195}(c)), during the initial phase of the prominence eruption. Then B1 separated into two ribbon-like structures moving away from each other (Fig.~\ref{fig:erupt_195}(d)). The 2nd episode of brightening (B2) appeared to be related to LS's eruption (Fig.~\ref{fig:erupt_195}(d)), and the 3rd episode (B3) took on the form of a two-ribbon flare (Fig.~\ref{fig:erupt_195}(f)), associated with the prominence eruption. Both B1 and B2 had an irregular, moss-like appearance initially and later became the footprints of some transient brightening loops labeled T1 and T2 in Fig.~\ref{fig:erupt_195}(e), see also the accompanying animation). Prior to the brightenings B1 and B2, two pairs of dimming regions were observed to be located at both sides of the prominence and the LS, respectively (marked by arrows in \fig{fig:erupt_195}(c) and (d)). Coronal dimmings are often interpreted as a mass deficit due to eruptions \citep[e.g.,][]{sh97,harrison03}. Dimmings in pair have only been  occasionally observed and was suggested to represent the feet of an eruptive flux rope \citep[e.g.,][]{thompson98,thompson00,webb00,liuc07}. 

\begin{figure}
\includegraphics[width=0.45\textwidth]{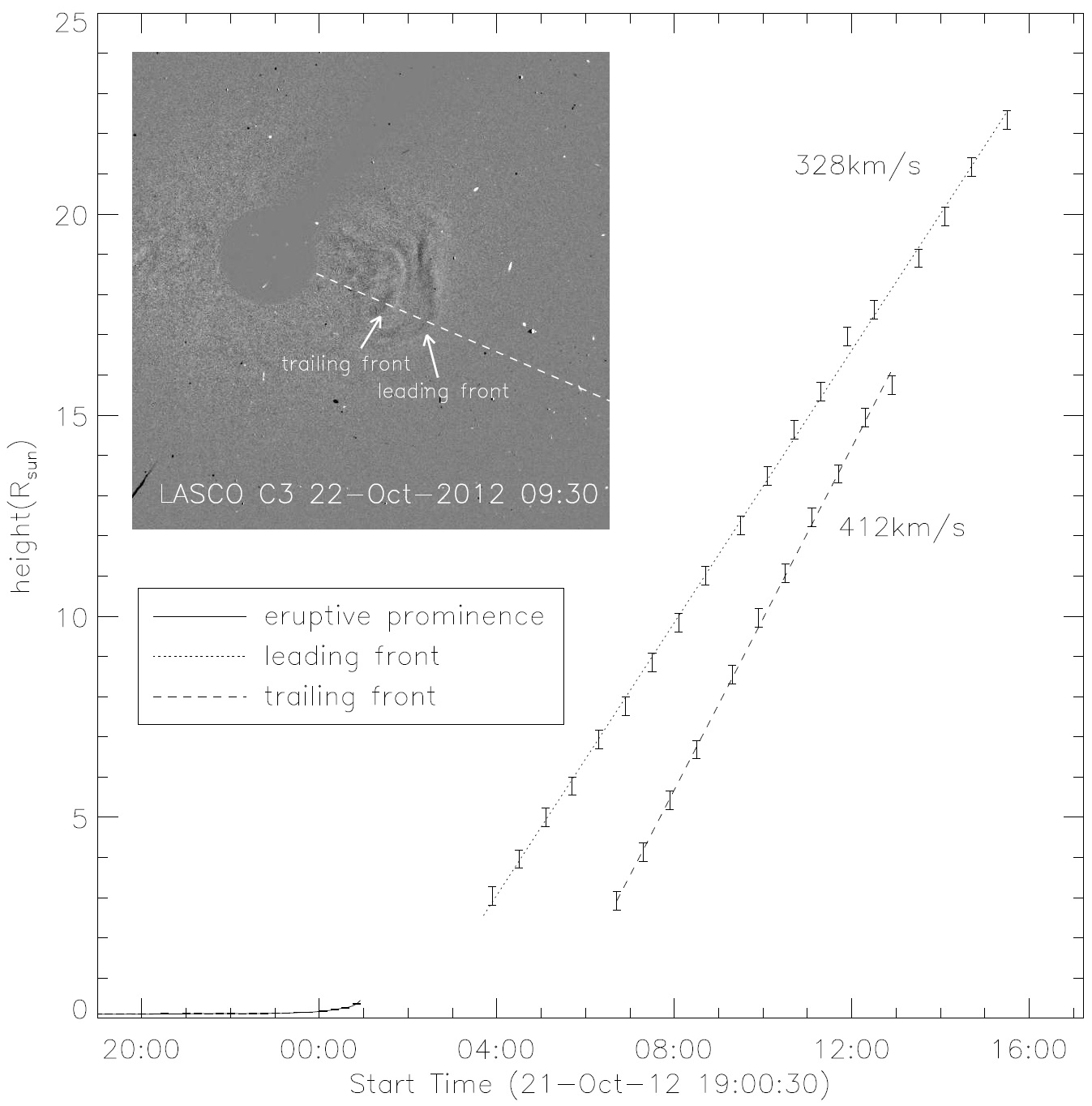}
\caption{Height-time evolution of the CME. The inset shows a LASCO/C3 white-light observation of the CME with two fronts (see the text for details).}
\label{fig:cme}
\end{figure}
The eruption results in two CME fronts as observed by LASCO/C3 (see the inset of \fig{fig:cme}). We may identify the source of the leading front with the LS and that of the trailing front with the prominence. Linear fitting of their height-time profiles yields that the leading front propagated at (328$\pm$2) km s$^{-1}$ in the plane of sky, slightly slower than the trailing front, which propagated at (412$\pm$4) km s$^{-1}$. Hence, the trailing front might eventually catch up with the leading front and interact with it \citep[e.g.,][]{shen12}.

\subsection{Pre-Eruption Dynamics}
\label{sec:pre-eruption}
To explore the physical mechanism of the eruption as described in \S\ref{sec:erupt}, we investigate the pre-eruption processes that might help make the prominence `ready' to erupt. What stands out is that within 2 days prior to its eruption the prominence was `fed' for at least three times by mini-prominences originally resting on the surface. This process is referred to as `flux feeding' hereafter. In this paper, we emphasize on the role of magnetic flux as far as `feeding' is concerned, although this process involves both magnetic and mass flux (see \S\ref{sec:instability}).

\subsubsection{Flux Feeding}
\label{sec:perturbations}
\begin{figure*}
\includegraphics[width=\textwidth]{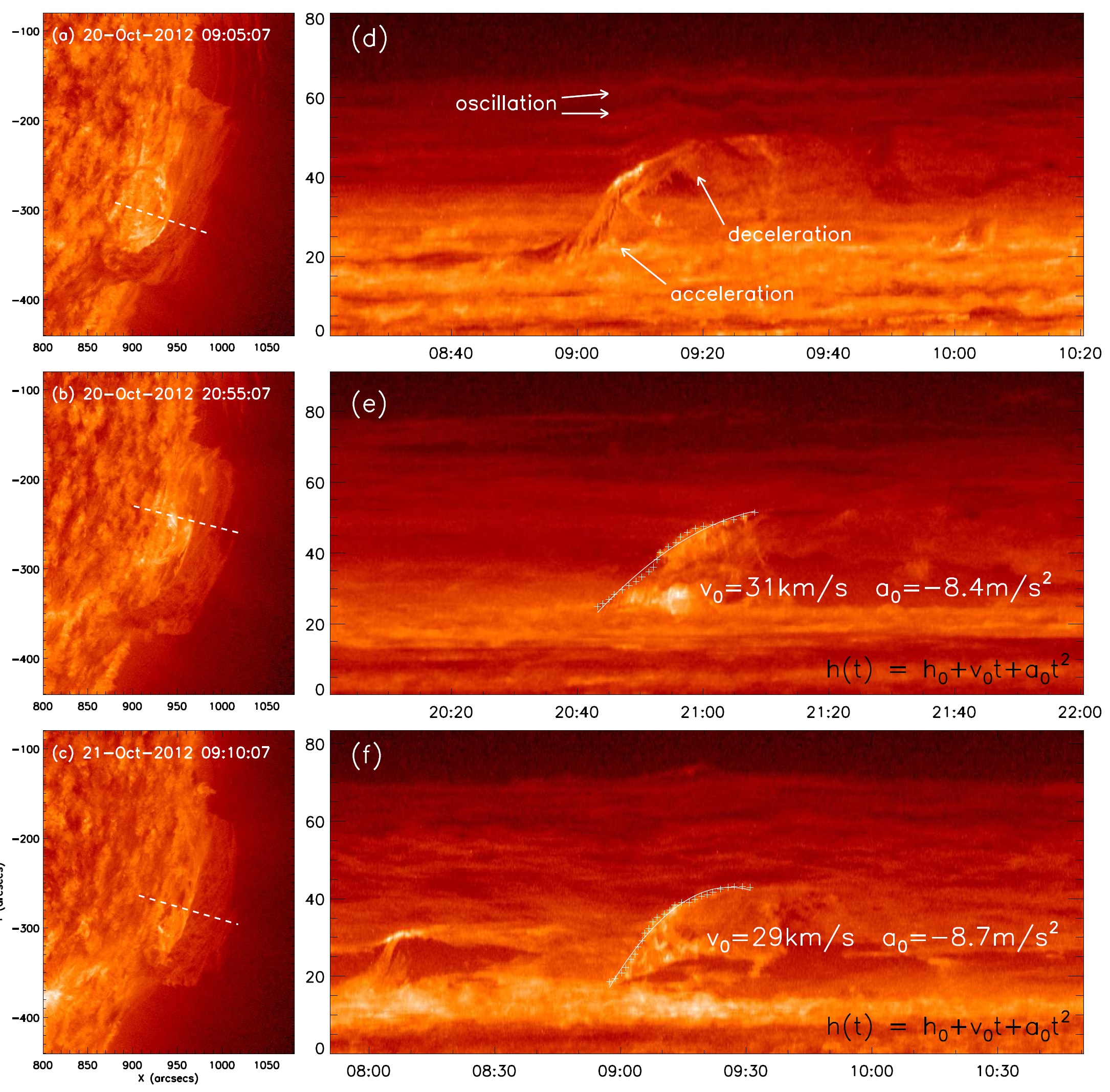}
\caption{Flux-feeding episodes. On each row, an individual episode is presented by the space-time stack plot in the right panel, which is made through the virtual slit marked by the dashed line in the left panel.}
\label{fig:total_slice}
\end{figure*}

The mini-prominences appear similar in emissivity as, but much smaller in spatial scales ($\sim 1/4$ in length and width) than, the target prominence in EUV images. In \fig{fig:total_slice} we study these feeding processes by placing a virtual slit along the rising direction of the mini-prominences, and present the resultant stack plots in a logarithm scale. During each feeding process, a miniature prominence rose upward apparently from the solar surface at a speed of tens of kilometers per second, interacted with, and eventually merged into, the target prominence. The interaction is characterized by an enhancement in brightness, and a decrease in speed, as the mini-prominences approached the target prominence. Each feeding process lasted for about half an hour.

\begin{figure*}
\includegraphics[width=\textwidth]{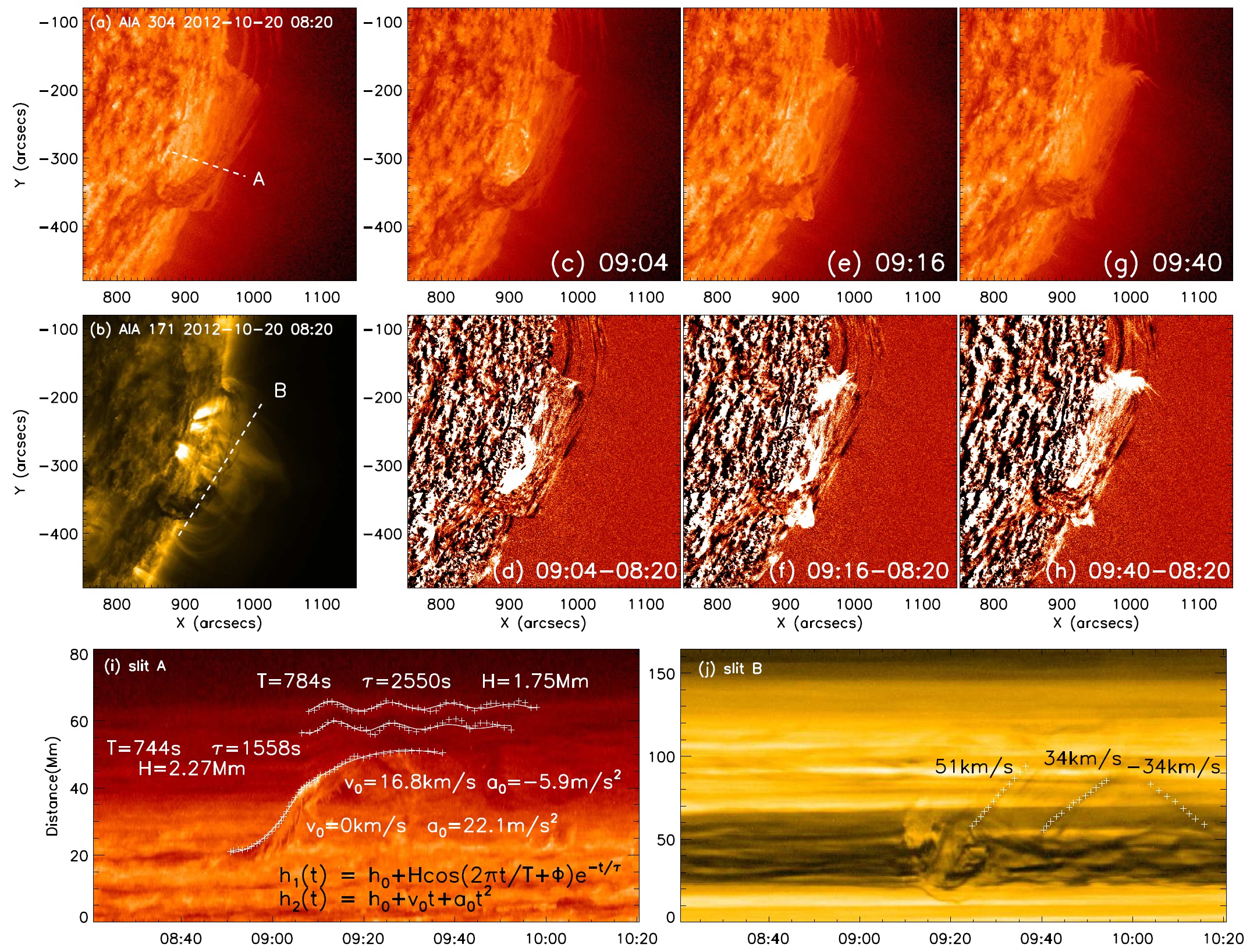}
\caption{First episode of flux feeding. Panels (a) and (b) show a 304~{\AA} and an 171~{\AA} image just before the feeding process, respectively, with two virtual slits A and B denoted by white dashed lines; Snapshots of the feeding process as observed in 304~{\AA} are shown in Panels (c), (e) and (g), and the corresponding base-difference images in Panels (d), (f) and (h); Panels (i) and (j) are the space-time stack plots obtained from the virtual slits in (a) and (b), respectively. Slit A is the same as that in \fig{fig:total_slice}(a). Fitting results on various features on the stack plots are also given (see the text for details). An animation of AIA 304 and 195~{\AA} images is available online.}
\label{fig:perturbation1}
\end{figure*}

The first feeding process, which took place at about 09:00 UT on 2012 October 20, is particularly interesting (see the animation accompanying Fig.~\ref{fig:perturbation1}). The upward-moving mini-prominence apparently drove the oscillation of two threads within the target prominence (\fig{fig:total_slice}(d)). Detailed analysis is shown in Figure \ref{fig:perturbation1}. The mini-prominence became visible as early as 08:20 UT, as a fibril-like structure. It started to rise at approximately 08:55 UT, with an acceleration of \accel{22$\pm$1} (\fig{fig:perturbation1}(i)). At about 09:05 UT, the upward moving turned into a deceleration of \accel{-5.9$\pm$0.6}, indicating an interaction with the prominence overhead. This compression process may result in the brightening of the mini-prominence from 09:05 UT onward, as well as the oscillation of the prominence threads at higher altitudes shortly after that time (see also \fig{fig:total_slice}(d)). The oscillation can be well fitted with a damped cosine function:
\begin{equation*}
h(t)=h_{0}+H\cos (\dfrac{2\pi}{T}t+\phi)\mathrm{e}^{-t/\tau},
\end{equation*}
where $H$, $T$, and $ \tau $ corresponds to the amplitude, period, and e-folding damping time, respectively. The two threads oscillated with essentially the same period, but the oscillation of the upper thread started slightly later, had a smaller amplitude, and decayed slower than the lower thread (Fig.~\ref{fig:perturbation1}(i)), suggesting that the oscillations were due to an upward-propagating wave which was excited by the interaction of the mini-prominence with the target prominence. The velocity amplitude was about 19 and 14 km~s$^{-1}$ for the lower and upper thread, respectively, significantly larger than that of small-amplitude oscillations (from 0.1 to several kilometers per second) that are apparently ever-present in prominences \citep{arregui12}. The present observation therefore provides an alternative cause for large-amplitude prominence oscillations (velocity amplitude $\gtrsim 20$ km~s$^{-1}$), which are relatively rare and have been suggested to be triggered by waves and disturbances produced by flares or jets \citep{tij09}. 

The eventual merge of the mini-prominence with the target prominence is characterized by knots of filament material moving along the prominence axis bi-directionally at tens of kilometers per second, reminiscent of counter-streaming flows in prominences \citep[e.g.,][]{zem98,lew03,lin05,ahn10,alexander13}. The space-time plot (\fig{fig:perturbation1}(j)) obtained from the virtual slit parallel to the prominence axis (Slit B in Figure \ref{fig:perturbation1}(b)) clearly shows that such horizontal motions became appreciable in 171~{\AA} images only after the merge at about 09:20 UT (also see the animation accompanying Fig.~\ref{fig:perturbation1}). However, counter-streaming flows are believed to be ubiquitous in prominences, despite that its cause remains unclear \citep[see][for a discussion]{chen14}. The fact that the horizontal motion excited by the disturbance from the rising mini-prominence well resembles counter-streaming flows suggests that such flows are dictated by the magnetic nature of prominences. It is known that the prominence field is dominantly horizontal and directed along the prominence axis \citep[e.g.,][]{leroy89}, which may explain the observed horizontal motions as well as the absence of vertical motions within the perturbed prominence. 

The hight-time profile of the mini-prominence is fitted with a piecewise parabolic function, with a uniform acceleration followed by deceleration. The fitting results are given in \fig{fig:perturbation1}(i). The second and third episodes of flux feeding are shown in the middle and bottom panels of \fig{fig:total_slice}, respectively. Unlike the first episode, there were no discernible oscillations resulting from the interaction, and the acceleration phase was less appreciable, so that both hight-time profiles can be well fitted with a uniform deceleration function with $v_0\approx 30$ km~s$^{-1}$ and $a\approx -8$ m s$^{-2}$. 

\subsubsection{Height-Time Evolution}
\label{subsec:height}

\begin{figure*}
\centering
\includegraphics[width=0.8\textwidth]{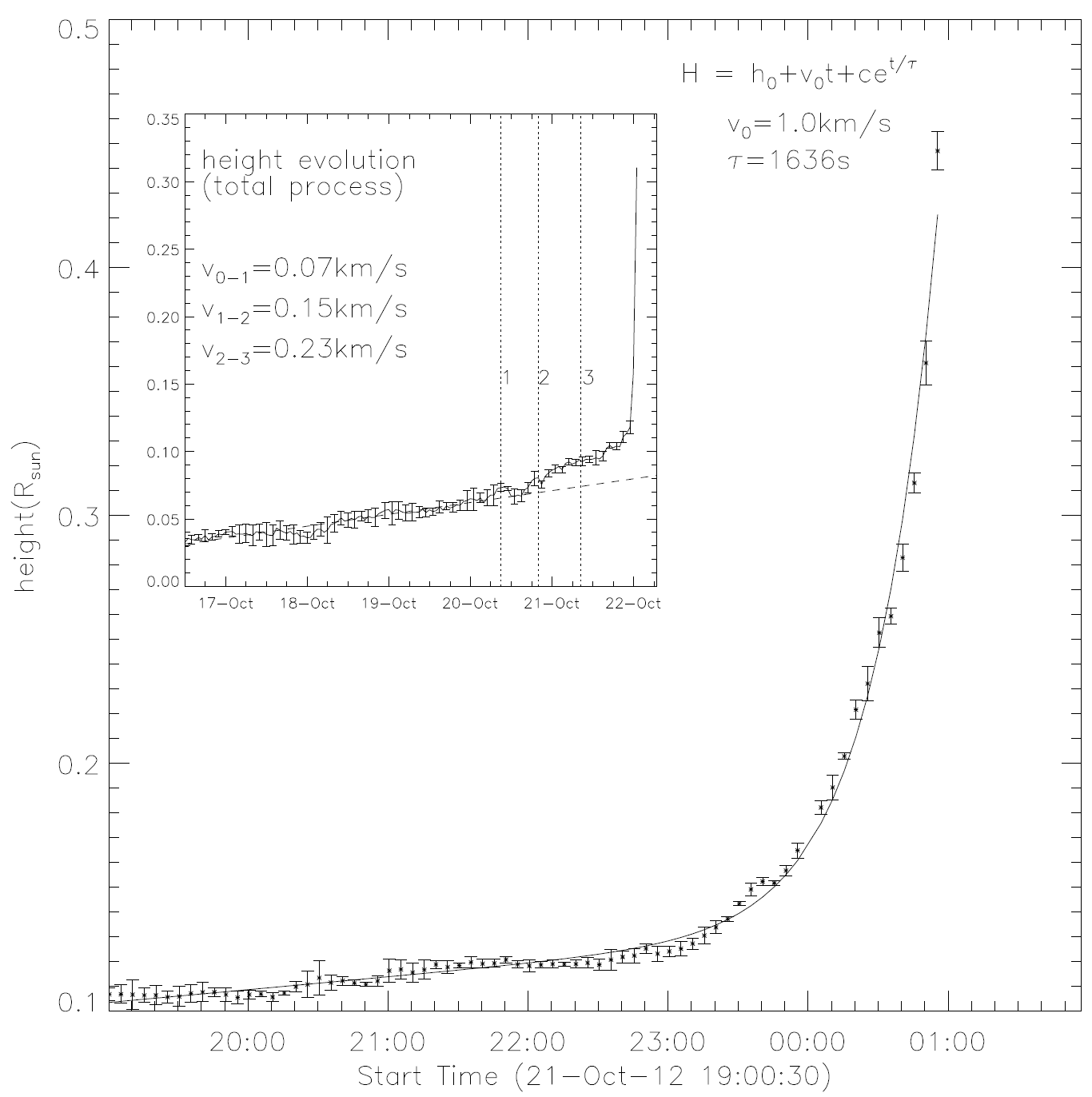}
\caption{Height-time evolution of the prominence. The leading edge of the prominence is measured with both SDO/AIA and STEREO-B/EUVI images until it left the AIA FOV. The profile after the third episode of flux feeding is fitted by an exponential function (shown as the solid curve). The inset shows the height-time profile starting on 2012 October 17 till October 22; three dotted lines mark the occurrences of flux feeding; the dashed line denotes the linear fitting result for the height-time evolution prior to any flux-feeding episodes.}
\label{fig:height}
\end{figure*}

That the flux-feeding processes affected the evolution of the prominence is evidenced by its height-time profile. With the \texttt{SCC\_MEASURE} procedure in SolarSoftWare, the `true' height, rather than projected height, of the prominence can be obtained. The inset of \fig{fig:height} shows the height-time profile of the prominence starting from 2012 October 17 till the prominence eruption on October 22. The vertical dashed lines mark the occurrences of the three episodes of flux feeding. The average speed till the occurrence of the 1st episode is $v_{0-1}\approx (0.071\pm0.002)$ km~s$^{-1}$. Similarly, $v_{1-2}\approx (0.150\pm0.070)$ km~s$^{-1}$ denotes the average speed between the 1st and 2nd episode, and $v_{2-3}\approx (0.227\pm0.028)$ km~s$^{-1}$ the average speed between the 2nd and 3rd episode. One can see that the average rising speed of the prominence was significantly enhanced with the flux-feeding processes, as compared with the long time interval before the occurrence of the 1st episode. This might be due to the magnetic-flux increase of the prominence, and therefore the strengthening of the outward magnetic pressure of the prominence field over the inward magnetic tension of the external field.

The eruptive process on 2012 October 22 can be well fitted by an exponential function with an initial height $h_0$ and velocity $v_0$:
\begin{equation*}
h(t)=h_0+v_0t+c\mathrm{e}^{t/\tau}.
\end{equation*}
The fitting yields that $v_0=(1.00\pm0.03)$ km~s$^{-1}$ and $\tau=(1636\pm14)$ s. Let the linear term equal to the exponential term, we are able to determine that the prominence eruption started at 0.154~$R_\odot$ at 23:49 UT on 2012 October 21.

\section{Discussion and Conclusion}
\label{sec:Discussion}

\subsection{Nature of Mini-Prominences} \label{sec:nature}
The flux-feeding process is reminiscent of the flux transfer within the double-decker filament as reported by \citet{liu12}. In the present study, the merge of the mini-prominences into the target prominence also feeds magnetic flux and current to the latter, resulting in an increasing speed of its quasi-static ascent, and eventually leading up to its unstableness. So, are the mini-prominences part of the lower branch of a double-decker filament? The main body of the lower branch could be lying beneath the photosphere so that only the upper branch was observed as the target prominence. If the lower branch emerges, then a double-deck configuration ensues. The Hinode observation that a flux rope emerges under a pre-existing filament \citep{okamoto08} might be such a case.  

\begin{figure*}
\includegraphics[width=\textwidth]{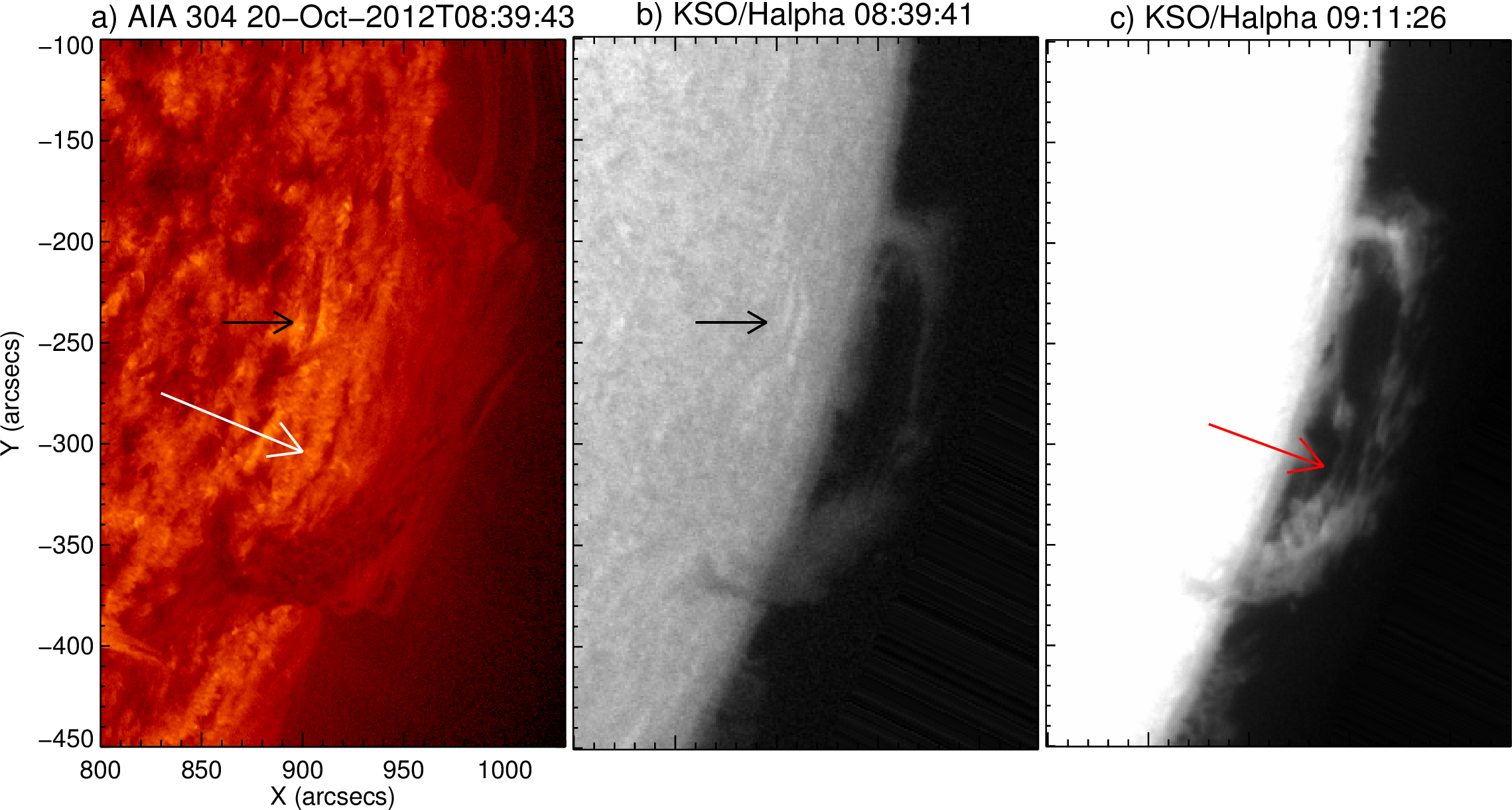}
\caption{Mini-prominence in AIA 304~{\AA} and H$\alpha$ as observed on 2012 October 20. The white arrow in Panel (a) marks the mini-prominence in 304~{\AA}, which is hardly visible in the H$\alpha$ image obtained at the same time (Panel (b)). Black arrows in Panels (a) and (b) mark a fibril structure. Panel (c) shows an H$\alpha$ image with the disk overexposed, which highlights the rising mini-prominence, as marked by a red arrow, on its way to merge into the target prominence.}
\label{fig:fibril}
\end{figure*}

Back to the present study, the mini-prominences were observed in AIA 304~{\AA} as thin elongated structures lying on the surface before rising upward to interact with the target prominence. In morphology, they are very similar to chromospheric fibrils, which cover most of the disk in H$\alpha$ line core. However, taking the first mini-prominence as an example (marked by a white arrow in Fig.~\ref{fig:fibril}(a)), one cannot easily find its counterpart in H$\alpha$ (Fig.~\ref{fig:fibril}(b)). In contrast, a slightly thicker fibril (marked by black arrows) at about $(900'',-230'')$ to the northeast of the mini-prominence can be seen in both 304~{\AA} and H$\alpha$. Hence, this could be due to the relatively poor resolution of the H$\alpha$ images, whose contrast are further plagued by the seeing conditions. As it rose above the limb, the mini-prominence can also be observed in H$\alpha$ (Fig.~\ref{fig:fibril}(c)), suggesting it is indeed of the same nature as fibrils. This is not surprising as filaments and fibrils are closely related in the sense that a prerequisite for a filament is a channel of chromospheric fibrils aligned with the polarity boundary, known as ``filament channel'' \citep{gaizauskas98,martin98}.

Arguably the tracer of the chromospheric field \citep{cs11, jing11}, fibrils can be regarded as small flux tubes. Apparently attracted to the target prominence, these rising fibrils must carry currents in the same direction, or, helicity of the same sign, as the prominence, in light of the MHD simulations of interactions between parallel flux tubes done by \citet{linton01}. However, there still exists a possibility that these fibrils actually belong to a flux rope lying beneath the surface, serving as the lower branch of a double-deck configuration.

It is worth commenting that the rising fibrils are distinct from buoyant plumes detected in off-limb observations \citep{berger10}, which are dark, bubble-like features in visible-light spectral bands, rising and inflating through the bright prominence emission with approximately constant speeds. In contrast to the plumes, the fibrils appear in emission above the limb and in absorption on the disk, same as the prominence in terms of emissivity; they rise from the surface and merge into the prominence with obvious deceleration, but no significant inflation. 

\subsection{Role of Instability} \label{sec:instability}
The low-lying forward S-shaped prominence rotated clockwise at the onset of the eruption, and it is not clear whether it was still kinked during the eruption. That the S-shape was apparently straightened due to the clockwise rotation implies a reduction of writhe, therefore excluding the helical kink instability as the trigger of the eruption \citep{torok10,liu12}. However, the exponential rise of the prominence and the lack of intense heating during the initial phase of the eruption suggests that a certain instability or loss of equilibrium may play an important role. Here we discuss three related mechanisms: a) flux imbalance, b) mass loading, and c) torus instability.  

Numerical studies have suggested that a flux rope could become unstable due to an increase of the axial flux, whose amount may possess a threshold for the existence of stable equilibria \citep{bobra08, sv09, su11}. The threshold appears to be only 10\%--20\% of the total flux in the region. For the quiescent prominence in question, a rather modest amount of flux transfer to it through the rising fibrils may be significant enough to reach the critical point. 

On the other hand, the rising fibrils also input mass into the target prominence. Mass loading could help hold down current-carrying flux, therefore raising the amount of free magnetic energy that can be stored in the pre-eruption configuration \citep{low03}. Thus, mass loading may also play a role in the present case, except that we do not see a significant increase in the darkness or thickness of the target prominence during the pre-eruption evolution, as reported in some cases \citep[e.g.,][]{kilper09, guo10, liu12}. However, both darkness and thickness could be modulated by the solar rotation: the apparent darkness of the prominence in EUV is expected to decrease as it rotated off the west limb, due to a `deeper' line-of-sight integration of EUV emission in the foreground; the thickness is affected by projection effect as the shape of the prominence is by no means symmetric along the line of sight. It is therefore difficult to determine quantitatively how much mass has been loaded as time progresses. 

The torus instability \citep{kt06, Torok2007a} sets in if a flux rope rises to a critical height \citep{liuk12} at which the overlying field declines with height at a sufficiently steep rate \citep{liuy08,aulanier10,oz10,fan10}, i.e., the decay index $n\equiv-d\log(B_h)/d\log(h)$ exceeds a critical value of 1.5, where $B_h$ is the horizontal component of the potential field external to the flux rope. However, the equilibrium of the system becomes unstable already when $n$ approaches $n_\text{crit}$. For example, \citet{da10} found that $n_\text{crit}$ typically falls in the range [1.1--1.3] for both circular and straight current channels. Here, we calculate the average decay index along the filament at different altitudes using the potential-field source-surface (PFSS) approximation \citep[][see the right panel of Fig.~\ref{fig:hmi}]{sd03}. The critical height $h_\text{crit}$ at $n_\text{crit}=1.3-1.5$ is about 0.15--0.27~$R_\odot$, where the filament became nominally torus unstable. The prominence indeed takes off at about 0.15~$R_\odot$, according to the exponential fit (\S\ref{subsec:height}). Hence, we conclude that the torus instability is the major mechanism in triggering the prominence eruption. In light of flux imbalance, the role of flux feeding is to force the prominence to seek for equilibrium at higher and higher altitudes, as evidenced by the enhanced slow-rise speed after each flux feeding episode. Consequently, the prominence reached the unstable height much earlier: at the average slow-rising speed of 0.07 km~s$^{-1}$ prior to the flux-feeding episodes (\S\ref{subsec:height}), the filament would have reached the critical height of 0.15~$R_\odot$ by 23:56 UT on October 29. In other words, this quiescent prominence might have been quite stable without flux feeding.

\subsection{Role of Reconnection} \label{sec:reconnection}
\begin{figure*}
\centering
\includegraphics[width=0.9\textwidth]{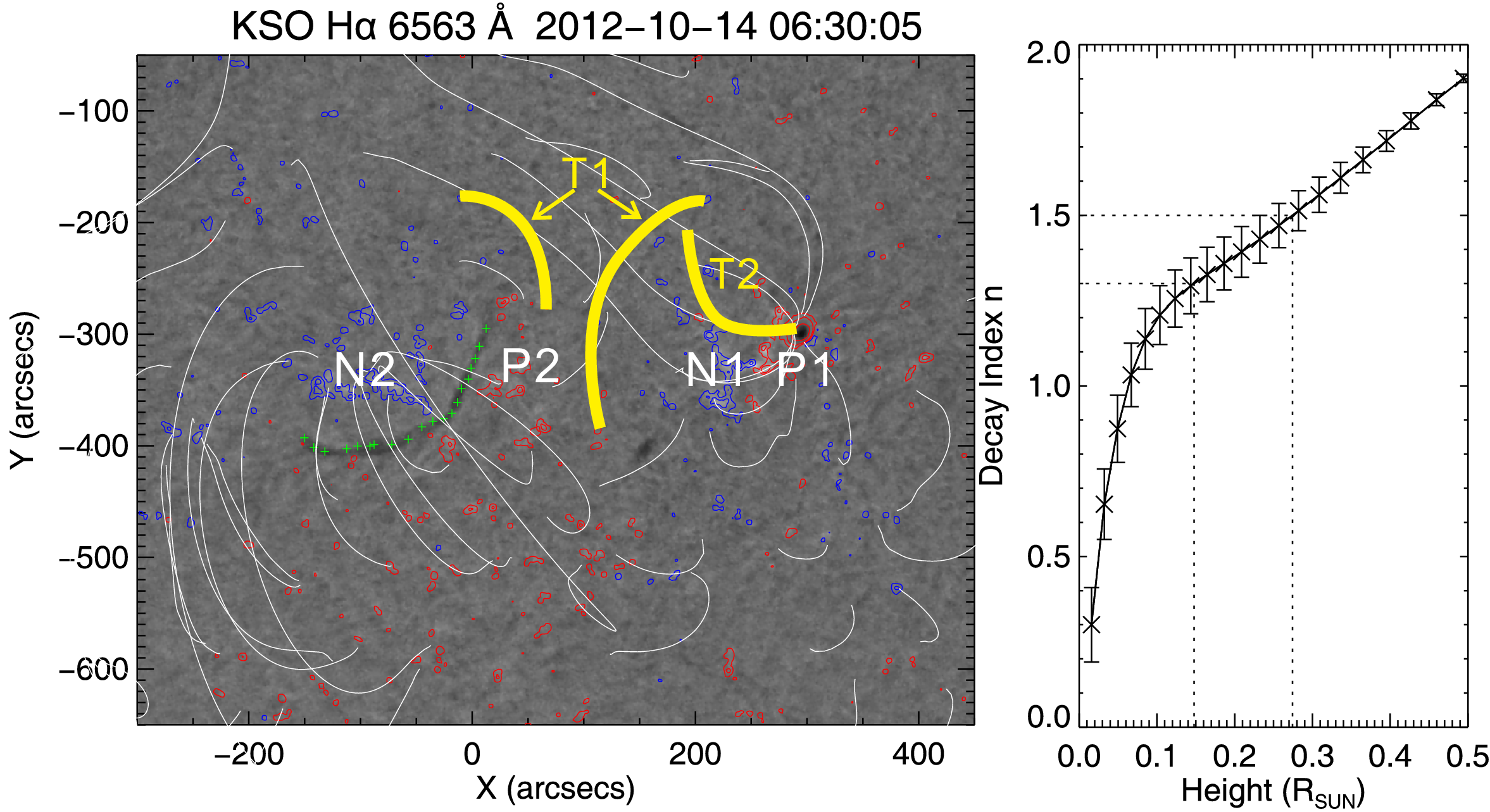}
\caption{Magnetic environment of the prominence in question as it crossed the central meridian on 2012 October 14. \emph{Left panel}: KSO H$\alpha$ image overlaid by the contours of the line-of-sight component of the photospheric magnetic field as obtained by SDO/HMI; contours levels are $\pm$50, $\pm$200, and $\pm$800 G, with red (blue) colors denoting positive (negative) polarities. Coronal field lines traced using the PFSS model are superimposed to demonstrate the large-scale magnetic connectivities.Four polarities of the quadrupolar configuration are labeled P1--N1 and P2--N2. The transient brightening loops T1 and T2 (Fig.~\ref{fig:erupt_195}(e)) that evolve from B1 and B2 are sketched with thick yellow curves. \emph{Right panel}: variation of the decay index $n$ with height, which is calculated using the PFSS model and averaged over the hand-picked points (green `+' symbols in the left panel) along the filament. The error bars reflect the standard deviation.}
\label{fig:hmi}
\end{figure*}

One can see that the prominence was embedded in a quadrupolar configuration (Fig.~\ref{fig:hmi}; the four polarities are labeled P1--N1 and P2--N2) by superimposing the line-of-sight component of the photospheric magnetic field upon the H$\alpha$ image taken on 2012 October 14 when the prominence was crossing the central meridian. A small bipolar active region located to the west of the filament was composed of a leading sunspot of positive polarity (P1) followed by diffused flux of negative polarity (N1). With the filament and the sunspot serving as landmarks, one can see that the LS as identified in EUV observations (Figs.~\ref{fig:orig} and \ref{fig:total_erupt}) must be connecting N1 and P1. During the eruptive process (\S\ref{sec:erupt}), successive surface brightenings in EUV (Fig.~\ref{fig:erupt_195}) first took place between the filament and the LS (B1), then in the active region on the west, associated with the eruption of the LS (B2), and finally on the east, associated with the prominence eruption (B3). All three brightening episodes, especially B1 and B3, were similar to two-ribbon flares in terms of both morphology and dynamics. B2 appeared to have only one ribbon (\fig{fig:erupt_195}(d)), but similar to B1 and B3 it had the moss-like appearance initially and later became the footprint of transient brighteing loops T2 below the erupting LS (\fig{fig:erupt_195}(e)). 

\begin{figure}
\includegraphics[width=0.45\textwidth]{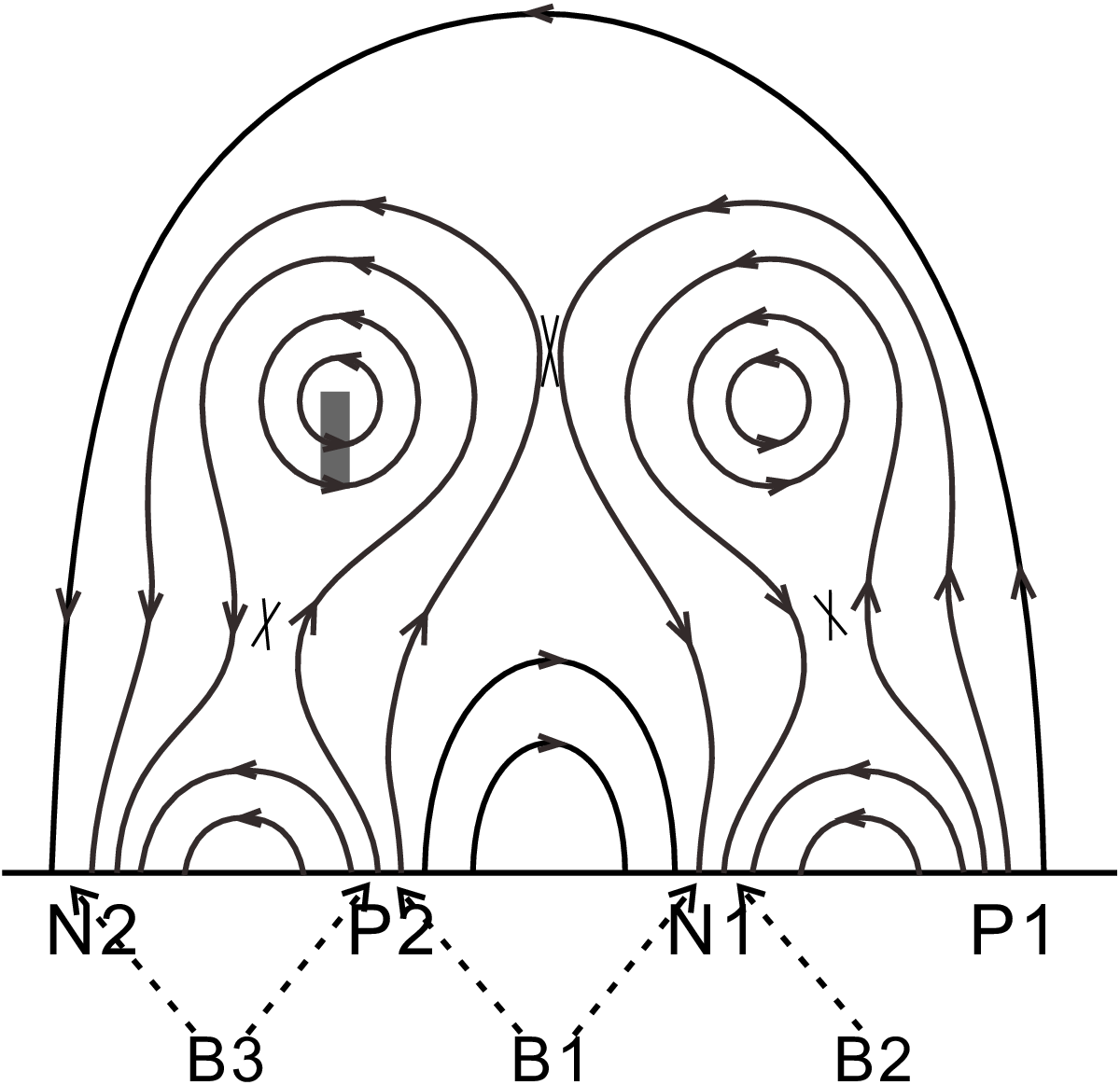}
\caption{Schematic of the quadrupolar magnetic field configuration in which the prominence is embedded. A gray slab indicates the body of the prominence. `X' symbols mark the locations of magnetic reconnections, which result in paired brightening ribbons on the surface. The observed brightenings, B1, B2, and B3, are marked by dashed arrows (see the text for details). }
\label{fig:cartoon}
\end{figure}

Based on these observations, we interpret the prominence eruption on 2012 October 22 in the framework of a schematic quadrupolar configuration (\fig{fig:cartoon}). Both the prominence and the LS are represented by a flux rope embedded in the two side-arcades. Within the two-day period prior to the eruption, multiple chromospheric fibrils rise upward and merge into the target prominence. The fibrils are apparently parallel to the target prominence and their interaction with the prominence results in horizontal flows along the prominence axis. We therefore speculate that primarily axial flux is ejected into the field of the prominence. With the accumulation of the axial flux, the prominence has to seek for equilibrium at higher heights. At certain point, it starts to interact with the flux rope embedded in the west side-arcade. The reconnection between the two flux ropes is evidenced by the first episode of two-ribbon brightening (B1) underneath the central arcade. The reconnection also cuts the `tethers' that hold down both flux ropes, leading to their rapid rise. Both rising flux ropes stretch their overlying fields and result in further reconnections underneath, which is evidenced by brightenings B2 and B3. It is remarkable that magnetic reconnection, as demonstrated by the surface brightenings, set in almost two hours after the eruption onset (\S\ref{subsec:height}). Thus, ideal instability  must dominate the initial phase of the eruption, though it was later coupled to reconnection to drive the eruption. 

We further conjecture that being held down by dense material causes the prominence eruption to progress initially on a slower pace than the LS in the west side-arcade, as evidenced by the fact that B2 precedes B3; but later on, the draining of the prominence material back to the surface (see the animation accompanying Fig.~\ref{fig:total_erupt}) may help the CME front resulting from the prominence eruption to catch up with the front caused by the erupting LS (see Fig.~\ref{fig:cme}). However, one must be aware of the limitation of this simplified scenario: despite deviating significantly from the potential field (Fig.~\ref{fig:hmi}), the observed brightening loops (T1) evolving from B1 do not connect B1's two ribbons, which cannot be explained by this 2D cartoon, but might be a reflection of the complex 3D nature of the reconnection between two flux ropes.

To conclude, we have described a new mechanism for a prominence to become torus unstable, i.e., chromospheric fibrils that carry the helicity of the same sign as the prominence could feed flux and current into the prominence, which results in the faster quasi-static ascent of the prominence, eventually leading up to its unstableness. We have also described a new paradigm of quadrupolar eruptions, i.e., two flux ropes embedded in the two side-arcades first interact to cut the constraining `tethers', and consequently erupt in close succession and proximity, effectively manifesting as a `twin' eruption \citep[e.g.][]{shen13}.

\acknowledgments The authors are grateful to the \sat{sdo}, \sat{stereo} and \sat{soho} consortium for the free access to the data. KSO H$\alpha$ data are provided through the Global H-alpha Network operated by New Jersey Institute of Technology. RL acknowledges the Thousand Young Talents Programme of China, NSFC 41222031 and NSF AGS-1153226. This work was also supported by NSFC 41131065 and 41121003, 973 key project 2011CB811403, CAS Key Research Program KZZD-EW-01-4, the fundamental research funds for the central universities WK2080000031.


\end{document}